\documentclass[showpacs,twocolumn,showkeys,amsmath,amssymb,pra,superscriptaddress,nofootinbib]{revtex4-1}

\usepackage{color}

\usepackage{graphicx}

\begin{document}

\title{Generating mesoscopic Bell states via collisions of distinguishable quantum bright solitons} 

\author{Bettina Gertjerenken}
\affiliation{Institut f\"ur Physik, Carl von Ossietzky Universit\"at, D-26111 Oldenburg, Germany}

\author{Thomas P.\ Billam }
\affiliation{Jack Dodd Center for Quantum Technology, Department of Physics, University of Otago, Dunedin 9016, New Zealand}

\author{Caroline L.\ Blackley}

\author{C.~Ruth {Le~Sueur}}
\affiliation{Joint Quantum Centre  (JQC) Durham--Newcastle, Department of Chemistry, Durham University, Durham DH1 3LE, United Kingdom}

\author{Lev Khaykovich}
\affiliation{Department of Physics, Bar-Ilan University, Ramat-Gan, 52900 Israel}

\author{Simon L.\ Cornish}

\author{Christoph Weiss}
\email{Christoph.Weiss@durham.ac.uk}
\affiliation{Joint Quantum Centre  (JQC) Durham--Newcastle, Department of Physics, Durham University, Durham DH1 3LE, United Kingdom}

\keywords{Bell state, matter-wave soliton, bright soliton, effective potential method, classical field method}
                  
\date{August 27, 2013}
 
\begin{abstract}
We investigate numerically the collisions of two distinguishable quantum
matter-wave bright solitons in a one-dimensional harmonic trap. We show that
such collisions can be used to generate mesoscopic Bell states which can
reliably be distinguished from statistical mixtures.  Calculation of the
relevant s-wave scattering lengths predicts that such states could potentially
be realized in quantum-degenerate mixtures of \textsuperscript{85}Rb and \textsuperscript{133}Cs.  In addition to
fully quantum simulations for two distinguishable two-particle solitons, we use a mean-field
description supplemented by a stochastic treatment of quantum fluctuations in
the soliton's center of mass: We demonstrate the validity of this approach by
comparison to a mathematically rigorous effective potential treatment of the quantum many-particle
problem.
\end{abstract} 
\pacs{03.75.Gg, 
 03.75.Lm, 
03.75.Mn, 
67.85.-d, 	
}
\maketitle 


Generating quantum entanglement between mesoscopic objects over mesoscopic
distances allows exploration of a fascinating ``middle-ground'' between quantum
and classical physics~\cite{HackermullerEtAl2004,Zurek2003}, and promises
significant advances in quantum-enhanced
interferometry~\cite{GiovannettiEtAl2004}.  The high degree of experimental
control offered by quantum-degenerate gases makes them an ideal platform with
which to explore such multi-particle entanglement \cite{EsteveNature2008,
RiedelNature2010}.  From a fundamental perspective, the creation of
\textit{maximally-entangled} many-particle Bell states in quantum-degenerate gases presents an intriguing
proposition. The generation of similar macroscopic Bell states of many photons is
an area of current theoretical and experimental research~\cite{StobinskaPRA2012,
IskahovPRL2012}. In addition to their inherent fundamental interest, such
states are of potential application as a resource in quantum information~\cite{IskahovPRL2012}.

Previously, the scattering of quantum bright matter-wave
solitons~\cite{KhaykovichEtAl2002,StreckerEtAl2002,BuljanEtAl2005,CornishEtAl2006,SachaEtAl2009,LewensteinMalomed2009,ErnstBrand2010,MartinRuostekoski2012,DamgaardHansenEtAl2012,CuevasEtAl2013}
in quasi-one-dimensional (1D) trapping geometries has been suggested as a way
to create mesoscopic entangled states in single-species Bose-Einstein
condensates (BECs)~\cite{WeissCastin2009,StreltsovEtAl2009b,LewensteinMalomed2009}.  In this Letter
we consider a dual-species BEC~\cite{McCarronEtAl2011,TakekoshiEtAl2012}, and
show that collisions of distinguishable quantum bright matter-wave solitons can
be used to generate mesoscopic Bell states~\cite{CsireApagyi2012}
(cf.~\cite{GarciamarchEtAl2011}),
\begin{equation}
\label{eq:Bell}
|\psi_{\rm Bell}\rangle\equiv\frac1{\sqrt{2}}\left(|{\rm A},{\rm B}\rangle+e^{i\alpha}|{\rm B},{\rm A}\rangle\right),
\end{equation}
where $|A,B\rangle$ ($|B,A\rangle$)  signifies that the BEC A is on the left
(right) and the BEC B is on the right (left). In particular, we show that a
favorable combination of inter- and intra-species s-wave scattering lengths
means that such states may be realized using \textsuperscript{85}Rb and
\textsuperscript{133}Cs mixtures. We
also show that the interference properties of these bright-soliton Bell states
distinguish them from statistical mixtures.  In contrast to the Bell ground
states associated with double-well potentials, our collisionally-generated Bell
states are robust to the presence of asymmetries. 
While distinguishable solitons are essential to produce Bell states, entanglement generation for solitons of the same species was investigated in~\cite{LewensteinMalomed2009}.

For our quasi-1D system, we consider an 
experimentally motivated harmonic confinement $\omega = 2\pi f$.
Mixtures of ultracold gases can be confined in a common optical trap with the same trap frequencies~\cite{SafronovaEtAl2006}, yielding
\begin{equation}
\omega = \frac{2\pi}T\;;\quad
\lambda_{\rm A} = \sqrt{\frac{\hbar}{m_{\rm A}\omega}}\;;\quad\lambda_{\rm B} = \sqrt{\frac{\hbar}{m_{\rm B}\omega}},
\end{equation}
where  $m_{\rm A}$ ($m_{\rm B}$) is the atomic mass of species A (B);
the interactions~$g=h f_{\perp} a$ are set by the scattering lengths~$a$ and the perpendicular trapping-frequency, $f_{\perp}$ \cite{Olshanii1998}.

We use the Lieb-Liniger model~\cite{LiebLiniger1963} for two species with additional harmonic confinement 
\begin{eqnarray}
\label{eq:H}
\hat{H} = &-&\sum_{j=1}^{N_{\rm A}}\frac{\hbar^2}{2m_{\rm A}}\partial_{x_j}^2 + \sum_{j=1}^{N_{\rm A}-1}\sum_{n=j+1}^{N_{\rm A}}g_{\rm A}\delta\left(x_j-x_{n}\right) \nonumber\\
&-&\sum_{j=1}^{N_{\rm B}}\frac{\hbar^2}{2m_{\rm B}}\partial_{y_j}^2 + \sum_{j=1}^{N_{\rm B}-1}\sum_{n=j+1}^{N_{\rm B}}g_{\rm B}\delta\left(y_j-y_{n}\right) \nonumber\\
&+& \sum_{j=1}^{N_{\rm A}}\sum_{n=1}^{N_{\rm B}}g_{\rm AB}\delta\left(x_j-y_{n}\right) \nonumber\\
&+&\sum_{j=1}^{N_{\rm A}}\frac 12 m_{\rm A}\omega^2x_j^2+\sum_{j=1}^{N_{\rm B}}\frac 12 m_{\rm B}\omega^2y_j^2\,,
\end{eqnarray}
where $x_j$ ($y_j$) and $g_{\rm A}<0$ ($g_{\rm B}<0$) are the atomic coordinates and intra-species interactions of species A (B), and 
 $g_{\rm AB}\ge 0$ is the inter-species interaction.

We suggest to prepare the two solitons independently; for weak harmonic confinement a single soliton has the ground state energy (cf.~\cite{McGuire1964})
\begin{equation}
E_{\rm S}(N_{\rm S}) = -\frac1{24}\frac{m_{\rm S}g_{\rm S}^2}{\hbar^2}N_{\rm S}(N_{\rm S}^2-1)\,;\quad {\rm S} \in \{\rm A,B\}\,.
\end{equation}
Thus, our system has the total ground-state energy
\begin{equation}
E_0 = E_{\rm A}(N_{\rm A}) + E_{\rm B}(N_{\rm B}) \,.
\end{equation}
The total kinetic energy related to the center-of-mass momenta $\hbar K_{\rm S}$ ( ${\rm S}  \in \{\rm A,B\}$) of the two solitons reads 
\begin{equation}
E_{\rm kin} = \frac{\hbar^2 K_{\rm A}^2}{2N_{\rm A}m_{\rm A}} + \frac{\hbar^2 K_{\rm B}^2}{2N_{\rm B}m_{\rm B}}.
\end{equation}

We extend the low-energy regime investigated for single-species solitons in Refs.~\cite{WeissCastin2009,SachaEtAl2009,GertjerenkenEtAl2012} to two species: 
\begin{equation}
\nonumber
E_{\rm kin} < \min\{\Delta_{\rm A}, \Delta_{\rm B}\},\quad
\Delta_{\rm S} =  \left|E_{\rm S}(N_{\rm S}-1)- E_{\rm S}(N_{\rm S})\right|.
\end{equation}
In this energy regime, each of the quantum matter-wave bright solitons is energetically forbidden to break up into two or more parts.
Highly entangled states are characterized by a roughly 50:50 chance of finding the soliton~A (B) on the left/right combined with a left/right correlation close to one indicating that whenever soliton~A is on the one side, soliton~B is on the other:
\begin{align}
\label{eq:corr}
\gamma(\delta)&\equiv
\int_{\delta}^{\infty}dx_1\ldots\int_{\delta}^{\infty}dx_{N_{\rm A}}\int^{-\delta}_{-\infty}dy_1\ldots\int_{-\infty}^{-\delta}dy_{N_{\rm B}}|\Psi|^2\nonumber\\
&+\int^{-\delta}_{-\infty} dx_1\ldots\int^{-\delta}_{-\infty}dx_{N_{\rm A}}\int_{\delta}^{\infty}dy_1\ldots\int_{\delta}^{\infty}dy_{N_{\rm B}}|\Psi|^2\,,
\end{align}
where $\Psi= \Psi(x_1,\ldots,x_{N_{\rm A}},y_1,\ldots,y_{N_{\rm B}})$ is the many-particle wave function (normalized to one) and $\delta \ge 0$.  The correlation $\gamma(\delta)$ will serve as an indication of entanglement: Bell states (\ref{eq:Bell}) are characterized by $\gamma\simeq 1$ combined with a 50:50 chance to find soliton A either on one side or on the other.

 We begin by investigating entanglement-generating collisions of two distinguishable two-particle solitons (dimers). Discarding cases where the two solitons have distinct total masses $N_{\rm A}m_{\rm A}$ and  $N_{\rm B}m_{\rm B}$ (small differences in the total masses would introduce small asymmetries without changing the physics), leads to $m_{\rm A}=m_{\rm B}=m$, which corresponds to having two hyperfine states of the same species. To describe the collisions of the two dimers, we discretize the Hamiltonian~(\ref{eq:H}), yielding the Bose-Hubbard Hamiltonian (cf.~\cite{AltmannEtAl2003})
\begin{align}
\label{eq:BH}
\hat{H}_{\rm BH} = \sum_{\ell} &\left\{
\frac{U_{\rm A}}2\hat{a}^{\dag}_{\ell}\hat{a}^{\dag}_{\ell}\hat{a}^{\phantom{\dag}}_{\ell}\hat{a}^{\phantom{\dag}}_{\ell} + \frac{U_{\rm B}}2\hat{b}^{\dag}_{\ell}\hat{b}^{\dag}_{\ell}\hat{b}^{\phantom{\dag}}_{\ell}\hat{b}^{\phantom{\dag}}_{\ell}
+U_{\rm AB} \hat{a}^{\dag}_{\ell}\hat{a}^{\phantom{\dag}}_{\ell}\hat{b}^{{\dag}}_{\ell}\hat{b}^{\phantom{\dag}}_{\ell}\right.
\nonumber\\
 &-J \left(\hat{a}^{\dag}_{\ell} \hat{a}^{\phantom{\dag}}_{\ell+1}+\hat{a}^{\dag}_{\ell+1}\hat{a}^{\phantom{\dag}}_{\ell}
+\hat{b}^{\dag}_{\ell} \hat{b}^{\phantom{\dag}}_{\ell+1}+\hat{b}^{\dag}_{\ell+1}\hat{b}^{\phantom{\dag}}_{\ell}\right)
\nonumber\\
&\left.+C\ell^2\hat{a}^{\dag}_{\ell}\hat{a}^{\phantom{\dag}}_{\ell} + C
 \ell^2\hat{b}^{\dag}_{\ell}\hat{b}^{\phantom{\dag}}_{\ell}\right\}\,,
\end{align}
where $U_{\rm A}$, $U_{\rm B}$ and $U_{\rm AB}$ are the intra-species and inter-species interactions,  the hopping is given by $J\sim \hbar^2/(2md^2)$ for grid spacing $d\to 0$ and  $C\equiv 0.5m\omega^2d^2$.

\begin{figure}
\includegraphics[width=\linewidth]{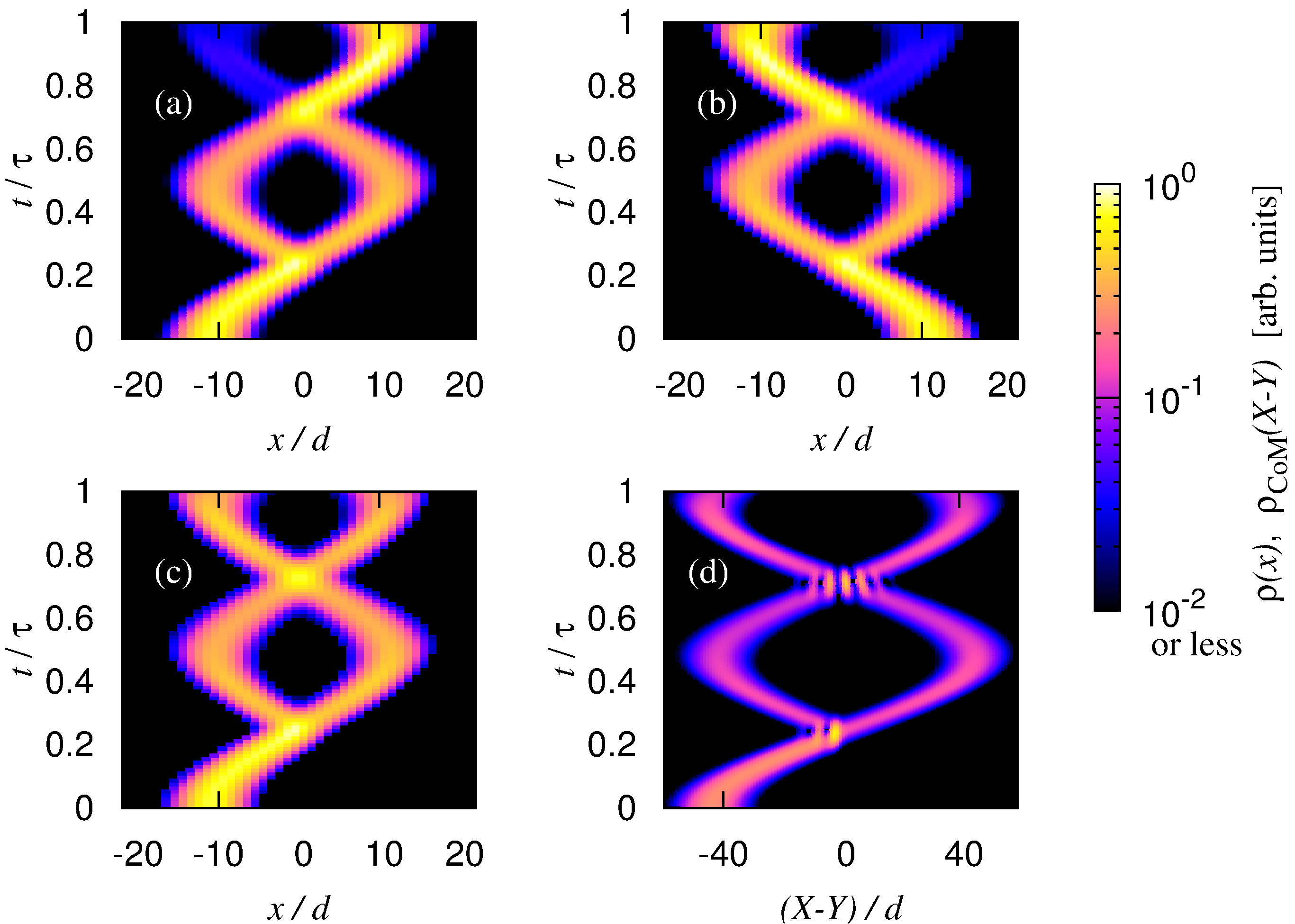}
\caption{\label{fig:quantum}(Color online) Collisions of two distinguishable dimers in the Bose-Hubbard Hamiltonian~(\ref{eq:BH}). \textbf{(a)} Single-particle density~$\varrho(x)$ of dimer A in a two-dimensional projection as a function of space and time ($\tau$ is the oscillation period without inter-species interaction, $U_{\rm A}=-3J$, $U_{\rm B}=-3J$, $U_{\rm AB}=J$ and $C=0.002J$). \textbf{(b)} Single-particle density of dimer B, parameters as in panel~(a).  \textbf{(c)} The same dimer as in panel~(a) but the wave function is numerically turned into a statistical mixture at $t=\tau/2$.   \textbf{(d)} Center-of-mass density $\varrho_{\rm CoM}(X-Y)$ if the inter-species interaction is switched off at $t=\tau/2$, all other parameters as in (a) and (b). The interference pattern near $t=0.7\tau$, combined with a high correlation~(\ref{eq:corr}) of $\gamma(d/2)\simeq 0.988$ near $t=0.5\tau$, indicates that a Bell state has been created.}
\end{figure}
Figure~\ref{fig:quantum} shows two-dimensional projections of the dynamics of two distinguishable dimers. The two dimers were numerically prepared in the ground state of two spatially separated harmonic oscillators via imaginary time-evolution~\cite{GlickCarr2011}. At time $t=0$ they were transferred into the same harmonic oscillator potential (without overlap). Subsequently, the time-evolution was calculated using the full Schr\"odinger equation corresponding to the Hamiltonian~(\ref{eq:BH}).
After the first collision, a measurement would reveal dimer A on the left and dimer B on the right or vice versa [the correlation~(\ref{eq:corr}) is $\gamma(d/2)\simeq 0.988$].

As the sizes of the dimers in panels Fig.~\ref{fig:quantum}~(a) and (b) are not too large compared to the oscillator length, after the second collision both dimers are more likely to be on the side opposite to their initial condition than at the same side (cf.\ the single soliton case~\cite{GertjerenkenEtAl2012}). This can be used to distinguish a pure quantum superposition from a statistical mixture [Fig.~\ref{fig:quantum}~(c)]. A more general approach extends the center-of-mass density~\cite{GertjerenkenWeiss2012} to two solitons [Fig.~\ref{fig:quantum}~(d)]: After switching off the interspecies interaction when the Bell-state has formed, one first measures the center of mass  $X$ and $Y$ of solitons A and B  and then plots the resulting density $\varrho_{\rm CoM}$ as a function of the difference $X-Y$. This works both for superpositions of plane waves $\exp[iKX]\exp[-iKY]+\exp[-iKX]\exp[iKY]$ with
\begin{equation}
\label{eq:CoM}
\varrho_{\rm CoM}(X-Y) \propto \{\cos[K(X-Y)]\}^2
\end{equation}
 and when the two wave packets recombine [Fig.~\ref{fig:quantum}~(d)]. Measuring a contrast close to one as in Eq.~(\ref{eq:CoM}) is possible as the center of mass can be measured with higher accuracy than the soliton width (cf.\ \cite{GertjerenkenWeiss2012}). As shown in Ref.~\cite{GertjerenkenWeiss2012} for the single-species case, CoM interferences do, in general, not correspond to interferences in single-particle densities which have been investigated, e.g., for distignuishable BECs in Ref.~\cite{CederbaumEtAl2007}.
\begin{figure}
\includegraphics[width=\linewidth]{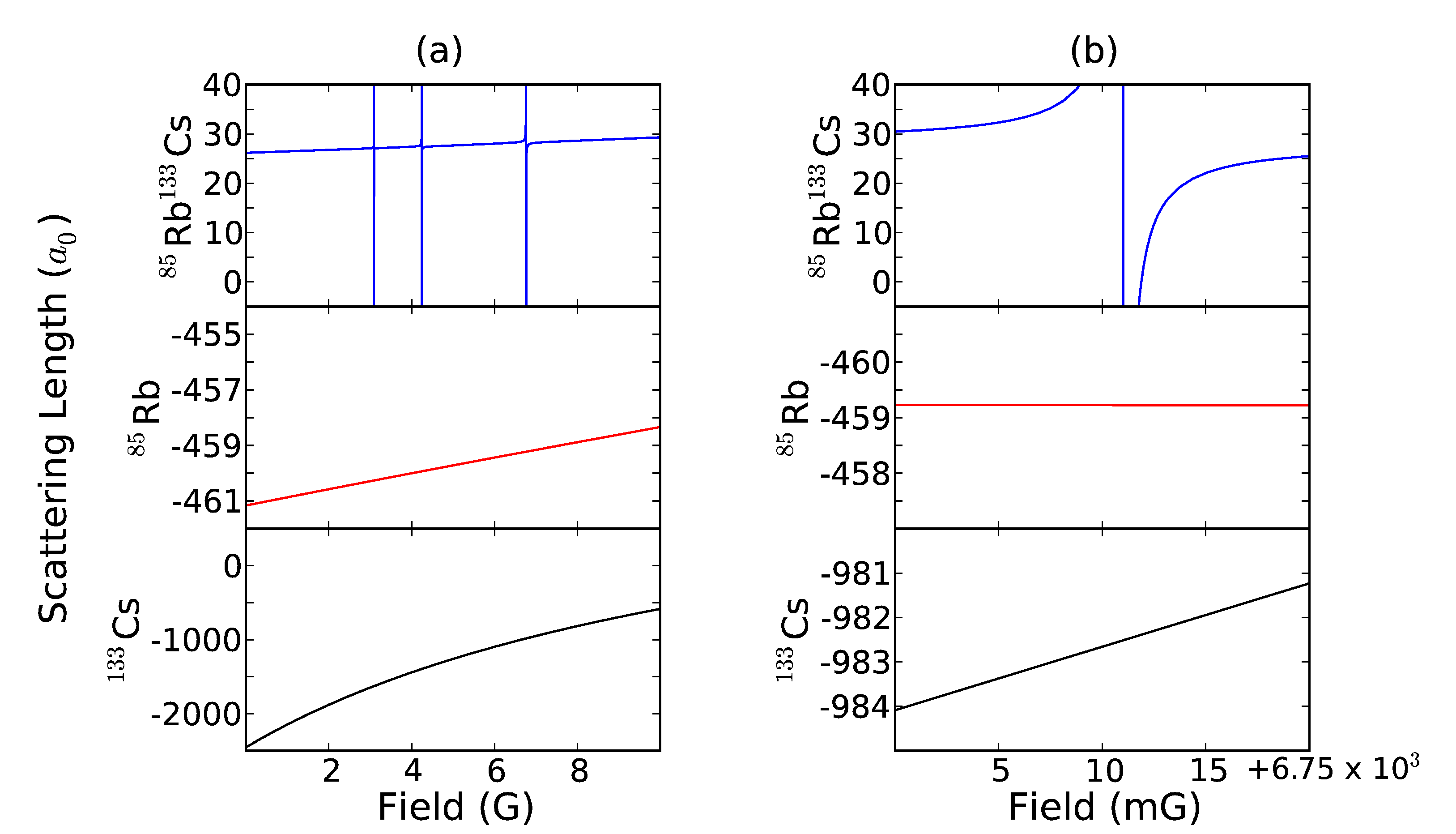}
\caption{\label{fig:rbcs}(Color online)
The s-wave scattering lengths for the ground state of \textsuperscript{85}Rb$^{133}$Cs, $^{85}$Rb and $^{133}$Cs respectively.  \textbf{(a)} Scattering lengths are calculated using a coupled-channels method \cite{TakekoshiEtAl2012} with a fully decoupled basis set at a collision energy of 1\,pK. The calculations are performed using the MOLSCAT program \cite{MOLSCAT} adapted to handle collisions in external fields \cite{GonzalezMartinezEtAl2007}. The RbCs potential is from \cite{TakekoshiEtAl2012}, Rb from \cite{StraussEtAl2010} and Cs from \cite{BerningerEtAl2011}. Resonances for $^{85}$Rb$^{133}$Cs are at 3.10\,G, 4.27\,G and 6.76\,G \cite{ChoEtAl2013}. \textbf{(b)} Zoom of (a).}
\end{figure}

To show that attractive intra-species interactions and repulsive, \textit{tunable}\/ inter-species interactions are experimentally feasible, we calculate the s-wave scattering lengths for  $^{85}$Rb$^{133}$Cs. The results displayed in Fig.~\ref{fig:rbcs}  shows a candidate inter-species Feshbach resonance at 6.76\,G suitable for our requirements~\cite{ChoEtAl2013}. 
For lower magnetic fields the magnetic field can be stabilized to up to 100\,$\mu$G~\cite{PasquiouEtAl2012}; shielding allows stabilization to 1\,mG below 10\,G. Although the masses of the atoms A and B now differ, we can still have two solitons of roughly the same total masses $N_{\rm A}m_{\rm A}$ and  $N_{\rm B}m_{\rm B}$ as in Fig.~\ref{fig:quantum}.

Behavior for larger particle numbers can be described by the Gross-Pitaevskii equation (GPE) (cf.~\cite{PuBigelow1998,Timmermans1998,OehbergSantos2001,HeEtAl2012})
\begin{align}
i\hbar\partial_t\varphi_{\rm A}(x,t) =& \left[-\frac{\hbar^2}{2m_{\rm A}}\partial_x^2+\frac{g_{\rm A}}2|\varphi_{\rm A}(x,t)|^2\right]\varphi_{\rm A}(x,t)\nonumber\\
&+\left[\frac12m_{\rm A}\omega^2x^2 +\frac{g_{\rm AB}}2|\varphi_{\rm B}(x,t)|^2\right]\varphi_{\rm A}(x,t)\nonumber\\
i\hbar\partial_t\varphi_{\rm B}(x,t) =& \left[-\frac{\hbar^2}{2m_{\rm B}}\partial_x^2+\frac{g_{\rm B}}2|\varphi_{\rm B}(x,t)|^2\right]\varphi_{\rm B}(x,t)\nonumber\\
&+\left[\frac12m_{\rm B}\omega^2x^2 +\frac{g_{\rm AB}}2|\varphi_{\rm A}(x,t)|^2\right]\varphi_{\rm B}(x,t)\;,\nonumber
\end{align}
where the single-particle density $|\varphi_{\rm S}(x,t)|^2$ is normalized to $N_{\rm S}$ ($ {\rm S} \in \{\rm A,B\}$). 

When hitting a barrier, the generic behavior of a mean-field bright soliton is to break into two parts; the fraction of the atoms transmitted decreases for increasing potential strength (cf.~\cite{MartinRuostekoski2012, CuevasEtAl2013}). An analogous behavior also occurs when two mean-field bright solitons hit each other as shown in the Supplemental
Material~\cite{GertjerenkenEtAl2013Sup}.

Low kinetic energies generate very different GPE dynamics.  For the case of
a single-species soliton incident upon a potential barrier one observes a sharp
stepwise jump in the GPE reflection coefficient as a function of barrier height~\cite{GertjerenkenEtAl2012,WangEtAl2012,DamgaardHansenEtAl2012}. In this case we previously~\cite{GertjerenkenEtAl2012} showed that this jump occurs in
regimes where, on the $N$-particle quantum level, the low kinetic energies
prevent the soliton from breaking into two (or more) smaller solitons,
and thus provides a useful GPE-level indicator for the formation of $N$-particle quantum superpositions.

Conjecturing that sharp stepwise jumps in the GPE reflection coefficient for distinguishable soliton collisions
may indicate Bell states, we investigate parameters yielding such jumps (cf.\  Supplemental Material~\cite{GertjerenkenEtAl2013Sup}).
To confirm that these jumps indicate Bell state formation, we
use the truncated-Wigner approximation (TWA), which describes quantum systems by
averaging over realizations of an appropriate classical field equation (in this
case, the GPE) with initial noise appropriate to either finite~\cite{BieniasEtAl2011} or zero temperatures~\cite{MartinRuostekoski2012}. While the GPE assumes both position and momentum are well
defined, this is not true for a single quantum particle of finite mass for
which, in general, both position and momentum involve quantum noise satisfying
the uncertainty relation. Our TWA calculations for the soliton
center-of-mass wave function use Gaussian probability distributions for both
(satisfying minimal uncertainty).

\begin{figure}
\includegraphics[width=\linewidth]{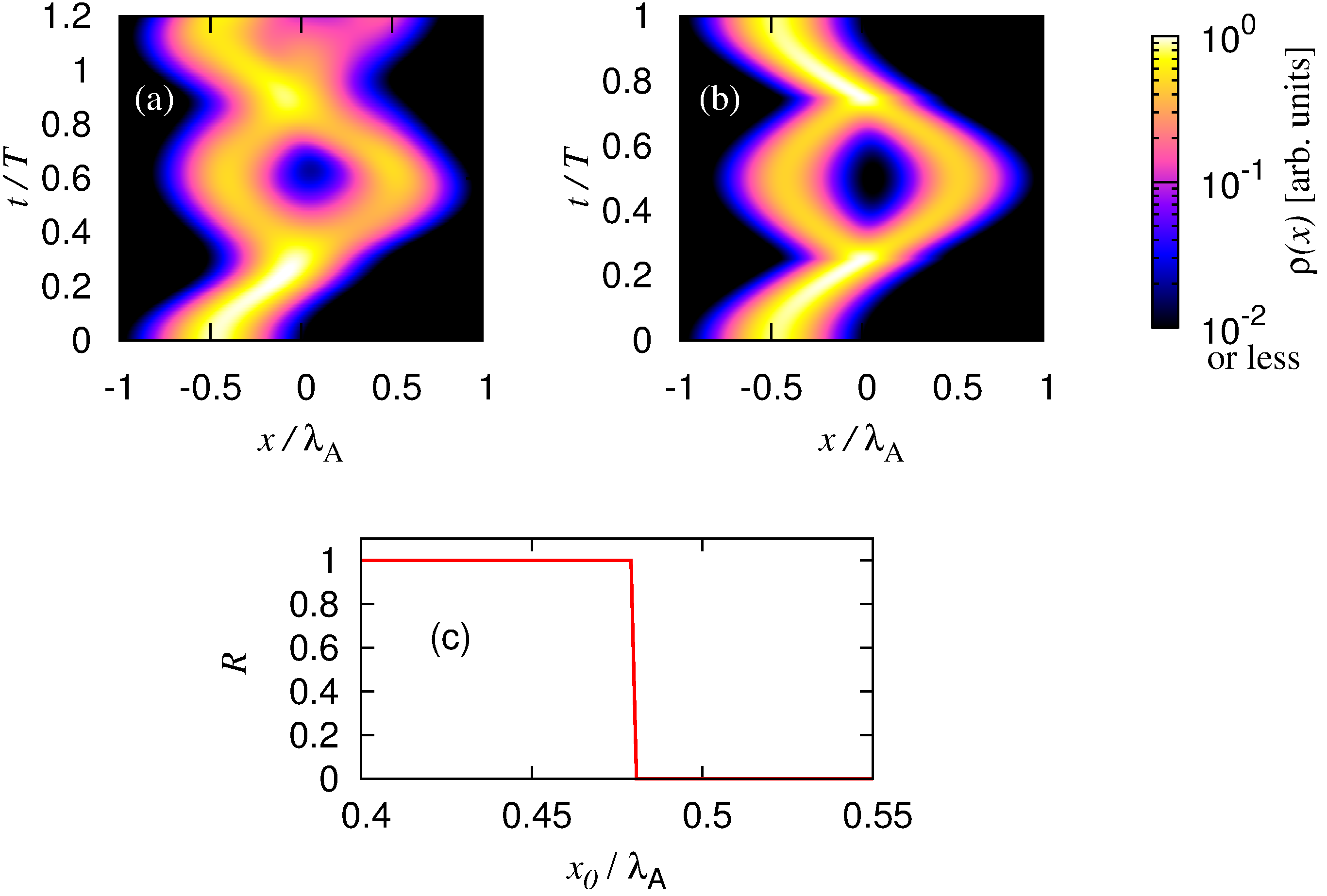}
\caption{\label{fig:TruncWigMot}(Color online) \textbf{(a)} Single-particle density for an 
$N$-particle quantum bright soliton hitting a narrow, heavy non-moving soliton,  computed using the effective potential approach (footnote~\ref{foot:N}). \textbf{(b)} As in (a) but using the TWA for the center of mass. The parameters are in the low kinetic energy regime such that the GPE predicts a sharp stepwise~\cite{GertjerenkenEtAl2012} behavior of reflection coefficient as a function of the initial displacement shown in panel~\textbf{(c)}; $U_0\simeq 12\hbar\omega$ (cf.\ footnote~\ref{foot:N}). 
}
\end{figure}

In order to demonstrate that the center-of-mass TWA is indeed a valid approach to describe the short-time behavior of mesoscopic quantum superpositions, Fig.~\ref{fig:TruncWigMot} starts with the case where a light soliton hits a heavy, non-moving soliton. In panel~(a), the rigorously proved~\cite{WeissCastin2012} effective potential approach~\cite{SachaEtAl2009,WeissCastin2009}\footnote{\label{foot:N}This approach replaces the $N$-particle Schr\"odinger equation by a single-particle Schr\"odinger equation for the center-of-mass coordinate. The effective potential is the convolution of the soliton with the potential seen by single particles~\cite{SachaEtAl2009,WeissCastin2009}, in our case the soliton B which is chosen to be a factor of $N_{\rm B}g_{\rm B}/(N_{\rm A}g_{\rm A})=10$ narrower than soliton A: $U_1 [\cosh(N_{\rm B}g_{\rm B}x)]^{-2}\simeq 2U_1/(N_{\rm B}g_{\rm B})\delta(x)$. If soliton A hits the narrow soliton B  [Fig.~\ref{fig:TruncWigMot}~(a)], the effective potential  $U_0 [\cosh(N_{\rm A}g_{\rm A}x)]^{-2}$ has the GPE-shape of the A-soliton with $U_0=N_{\rm A}U_1N_{\rm A}g_{\rm A}/(N_{\rm B}g_{\rm B})= 10U_1$, for $N_A=100$.} demonstrates the emergence of a Schr\"odinger-cat state when the GPE predicts the stepwise behavior of the reflection coefficient explained in Refs.~\cite{GertjerenkenEtAl2012,WangEtAl2012,DamgaardHansenEtAl2012} [Fig.~\ref{fig:TruncWigMot}~(c)].

In panel (b), we use the TWA to average over the analytic approximation for the classical-particle-like behavior of the GPE-soliton~\cite{MartinEtAl2008b}. This leads to a good qualitative agreement with the $N$-particle predictions in panel (a) up to the time where both parts of the wave function {recombine} and quantum interference becomes important. 

On the $N$-particle level, the low kinetic energies are important for the soliton not to be able to break into two (or more) smaller solitons. While GPE-solitons can, during a collision, lose a small fraction of particles, for low kinetic energies this effect becomes negligible [24]. Thus, the sharp stepwise behavior shown in Fig.~\ref{fig:TruncWigMot}~(c) leads to a behavior very close to the true $N$-particle quantum case.

\begin{figure}
\includegraphics[width=\linewidth]{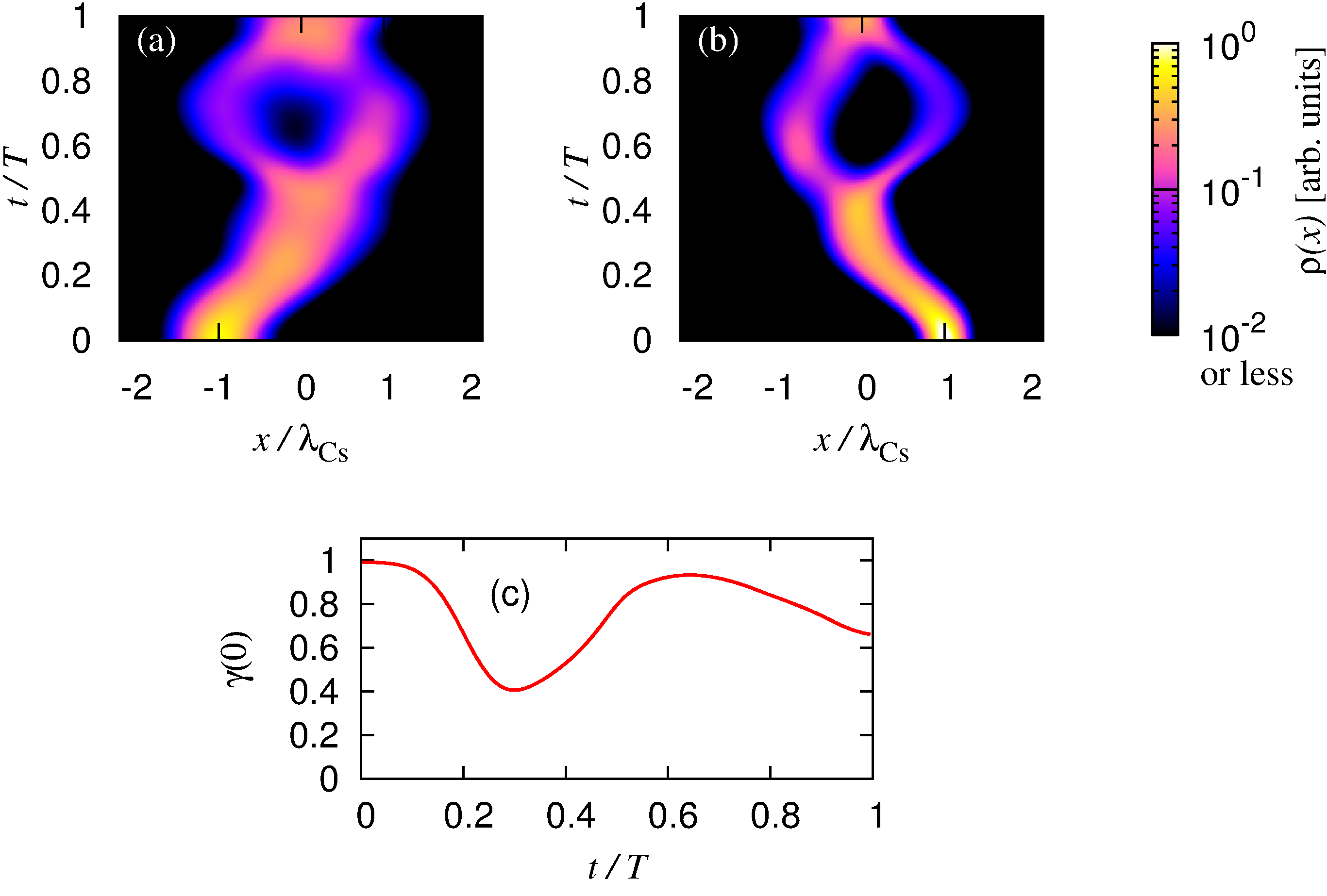}
\caption{\label{fig:TruncWigBin}(Color online)
TWA for the center of mass in the low-kinetic-energy regime applied to the two-species GPE. The single particle density for the  \textsuperscript{133}Cs-soliton is displayed in panel (a), for  \textsuperscript{85}Rb in panel (b). \textbf{d)} This leads to a correlation~(\ref{eq:corr}) close to one, thus indicating a Bell state. Parameters: $a_{\rm Cs}=-982.5a_0$, $a_{\rm Rb}=-459.2 a_0$, $f= 1$Hz, $f_{\perp}=70$Hz,
$N_{\rm Cs} \simeq  12$,
$N_{\rm Rb}\simeq 19$ (corresponding to $N_{\rm Rb}m_{\rm Rb}\approx N_{\rm Cs}m_{\rm Cs}$, thus avoiding center-of-mass movement),
$a_{\rm RbCs}=63.6 a_0$, and the initial
displacement from trap center: $- 8.7 \rm \mu m$ for \textsuperscript{133}Cs and $+ 8.7 \rm \mu m$ for  \textsuperscript{85}Rb.}
\end{figure}
In order to observe Bell states, we investigate two distinguishable bright solitons of similar mass at low kinetic energy  ($E_{\rm kin}/|E_0| = 0.182$).
 Applying the TWA for the center-of-mass wave functions of both solitons leads to the single-particle densities displayed in Fig.~\ref{fig:TruncWigBin}~(a) and (b). The low kinetic energies indicate that the feature shown in those single-particle densities  near $t\approx 0.6T$ should indeed be a Bell state. The value of the correlation function close to one [Fig.~\ref{fig:TruncWigBin}~(c)] shows that we indeed have found a Bell state. While the TWA is no longer valid as soon as both parts of the wave-function overlap, a full quantum mechanical calculation would also lead to a decrease of the correlation in Fig.~\ref{fig:TruncWigBin} (c)  on this time-scale.

To conclude, based on predictions made on the level of many-particle quantum
calculations (using the Lieb-Liniger model), we demonstrated numerically that
mesoscopic Bell states can be generated by colliding two distinguishable
quantum matter-wave bright solitons.  In experiment, the formation of these
states could be confirmed by switching off the inter-species interaction once
the Bell state has formed and then measuring the interference fringes in the
combined center-of-mass density (\ref{eq:CoM}) [see Fig.~\ref{fig:quantum}], revealing the presence of
quantum superposition. Finally, we have shown that matter-wave bright solitons
in \textsuperscript{85}Rb-\textsuperscript{133}Cs mixtures are a promising candidate system for experimental
realization of mesoscopic Bell states, presenting an intriguing target for
future experimental investigations.

\acknowledgments

We thank C.~S.~Adams, S.~A.~Gardiner, J.\ L.\ Helm, J.~M. Hutson and M.~P.~K{\"o}ppinger for discussions. We thank the \textit{Studienstiftung des deutschen Volkes} (B.G.), the \textit{Heinz Neum\"uller Stiftung} (B.G.), the Marsden Fund of New Zealand (contract UOO162) and the Royal Society of New Zealand (contract UOO004) (T.P.B), the Faculty of Science at Durham University (C.L.B.), the EOARD (Grant FA8655-10-1-3033, C.R.LS) and
 the UK EPSRC (Grant No.\ EP/G05
6781/1, C.W.) for funding.

\begin{appendix}

\section{Supplemental Material: Collision behavior of higher-kinetic-energy Gross-Pitaevskii bright solitons}

The mean-field approach via the Gross-Pitaevskii equation (GPE)~\cite{PethickSmith2008} provides physical insight into the behavior of bright solitons:
\begin{align}
i\hbar {\partial}_t\varphi(x, t) = &-\frac{\hbar^2}{2m}\partial_x^2 \varphi(x, t) +V_{\rm ext}(x) \varphi(x, t)\nonumber \\ &+(N-1)g_{1 \rm D}|\varphi(x, t)|^2 \varphi(x, t)\;,
\nonumber
\end{align}
where $m$ is the mass of one atom, $N$ is the number of atoms, $V_{\rm ext}(x)$ is the external potential and $g_{1 \rm D}$ quantifies the (contact-)interaction between two particles; the single-particle density $|\phi(x,t)|^2$ is normalized to one. 
 Without a scattering potential [$V_{\rm ext}(x)=0$], exact solutions for GPE-solitons exist~\cite{PethickSmith2008}
\begin{equation}
\varphi(x,0) =\sqrt{\frac{2\mu}{(N-1)g_{\rm 1D}}}\frac{e^{im\mu x/\hbar-i(\mu-mu^2/2) t/\hbar}}{\cosh\left[\sqrt{\frac{2m|\mu|}{\hbar^2}}({x-x_0 -ut})\right]},
\nonumber
\end{equation}
here $u$ is the  velocity and $x_0$ the initial position; normalizing $|\varphi(x,0)|^2$ to one yields (cf.~\cite{CastinHerzog2001})
\begin{equation}
\nonumber
\mu = -\frac 18\frac{mg_{\rm 1D}^2}{\hbar^2}(N-1)^2\;.
\end{equation}

In the regime of high kinetic energies~\cite{HolmerEtAl2007b,GertjerenkenEtAl2012}, 
\[
E_{\rm kin}\gg E_0,
\] where 
\[E_{\rm kin}=N\hbar^2k^2/(2m)\] and 
\[
E_0=-{mg_{\rm 1D}^2}N(N^2-1)/(24\hbar^2),\]
 scattering a soliton from a narrow barrier behaves essentially like scattering a single particle of such a barrier. For high enough kinetic energies, a bright soliton scattering off a delta-function barrier has the same transmission/reflection behavior~\cite{HolmerEtAl2007b} as the textbook example of a single particle scattered off such a potential~\cite{Fluegge1990}. 

There is, however, an important difference between the single particle case and high-energy bright solitons: For single particles, such a transmission/reflection behavior can only be measured by repeating the experiment often (the particle will always be measured either on one side or on the other side of the delta-function barrier). For a bright soliton, a transmission larger than 0 and lower than one leads to a (in an ideal experiment reproducible) fraction of the atoms being found on either side in \textit{each single experiment}.

In order to show that this ``classical'' breaking-into-two-parts  behavior also occurs for two collisions of two high-energy distinguishable bright solitons, we use the GPE (cf.~\cite{PuBigelow1998,Timmermans1998,OehbergSantos2001,HeEtAl2012})
\begin{align}
i\hbar\partial_t\varphi_{\rm A}(x,t) =& \left[-\frac{\hbar^2}{2m_{\rm A}}\partial_x^2+\frac{g_{\rm A}}2|\varphi_{\rm A}(x,t)|^2\right]\varphi_{\rm A}(x,t)\nonumber\\
&+\left[\frac12m_{\rm A}\omega^2x^2 +\frac{g_{\rm AB}}2|\varphi_{\rm B}(x,t)|^2\right]\varphi_{\rm A}(x,t)\nonumber\\
i\hbar\partial_t\varphi_{\rm B}(x,t) =& \left[-\frac{\hbar^2}{2m_{\rm B}}\partial_x^2+\frac{g_{\rm B}}2|\varphi_{\rm B}(x,t)|^2\right]\varphi_{\rm B}(x,t)\nonumber\\
&+\left[\frac12m_{\rm B}\omega^2x^2 +\frac{g_{\rm AB}}2|\varphi_{\rm A}(x,t)|^2\right]\varphi_{\rm B}(x,t)\;,\nonumber
\end{align}
where the single-particle density $|\varphi_{\rm S}(x,t)|^2$ is normalized to $N_{\rm S}$ ($ {\rm S} \in \{\rm A,B\}$),  $g_{\rm A}<0$ ($g_{\rm B}<0$) the intra-species interactions of species A (B), and 
 $g_{\rm AB}\ge 0$ is the inter-species interaction.

\begin{figure}
\includegraphics[width=\linewidth]{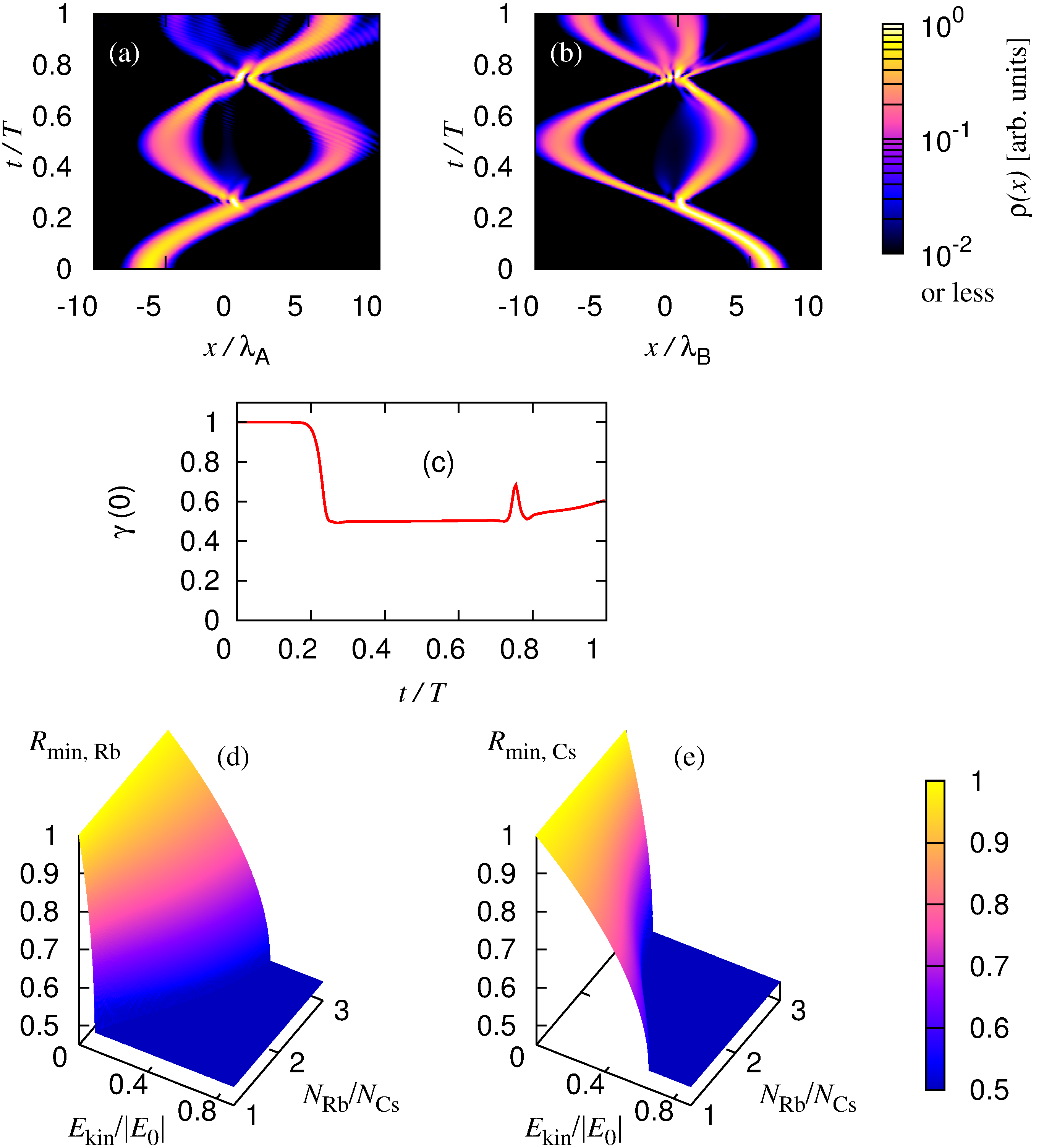}
\caption{\label{fig:jumps} \textbf{(a)} Single-particle density for a GPE-soliton of \textsuperscript{133}Cs atoms splits at high {kinetic} energies when hitting the soliton made of \textsuperscript{85}Rb atoms. {Parameters are} the \textit{s}-wave scattering lengths $a_{\rm Cs}=-982.5a_0$, $a_{\rm Rb}=-459.2 a_0$, $a_{\rm RbCs}=8351 a_0$, axial frequency $f= 1$Hz and radial frequency
 $f_{\perp}= 16$Hz, as well as particle numbers
$N_{\rm Cs} \simeq  17$, 
$N_{\rm Rb} \simeq 17$. {The starting displacement from the} trap center: {$\pm 55.2 \rm\mu m$} {\textbf{(b)} Single-particle density for the soliton made of \textsuperscript{85}Rb atoms for the simulation depicted in panel (a).}   \textbf{(c)} Although the single-particle densities of panels (a) and (b) look similar to what Bell-state would look like, at $t\approx T/2$ we have a low correlation~(\ref{eq:corrS}) and therefore no Bell state. 
\textbf{(d)} and \textbf{(e)} show lower bounds on $R_{\min}$ [Eq.~(\ref{eq:Rmin}) cf.~\cite{GertjerenkenEtAl2012}].
}
\end{figure}

Figure~\ref{fig:jumps} shows typical GPE dynamics in the regime where the kinetic
energy $E_{\rm kin}$ is large compared to the ground state
energy $|E_0|$ ($E_{\rm kin}/|E_0| = 84.9$). Here, both solitons split such that close
to 50\% of the particles of each species are on each side of the center of the
trap for times near $t = T/2$, and the correlation function
\begin{align}
\label{eq:corrS}
\gamma(\delta)&\equiv
\int_{\delta}^{\infty}dx_1\ldots\int_{\delta}^{\infty}dx_{N_{\rm A}}\int^{-\delta}_{-\infty}dy_1\ldots\int_{-\infty}^{-\delta}dy_{N_{\rm B}}|\Psi|^2\nonumber\\
&+\int^{-\delta}_{-\infty} dx_1\ldots\int^{-\delta}_{-\infty}dx_{N_{\rm A}}\int_{\delta}^{\infty}dy_1\ldots\int_{\delta}^{\infty}dy_{N_{\rm B}}|\Psi|^2\,,
\end{align}
 depicted in Fig.~\ref{fig:jumps}
(c) shows no signs of Bell states (which would lead to correlations close to one).

We extend the exact lower bounds on the
reflection coefficient $R$ for a single soliton scattered off a barrier~\cite{GertjerenkenEtAl2012} to
the case of two distinguishable solitons [Fig.~\ref{fig:jumps}~(d) and (e)], where
\begin{equation}
\label{eq:Rmin}
R_{\rm min}\equiv \min\{R\ge 0.5\}\;.
\end{equation}
Panels (d) and (e) of Fig.~\ref{fig:jumps} show that using roughly equal total masses $N_{\rm A}m_{\rm A}$ and  $N_{\rm B}m_{\rm B}$, $N_{\rm Rb}\approx 1.6 N_{\rm Cs}$ is indeed a good idea for entanglement generation: at approximately this ratio do jumps occur in both reflection
coefficients for kinetic energies corresponding to reasonable timescales
(without a trap) or initial displacements (in the presence of a trap).

Only by focusing on low kinetic energies, two distinguishable bright solitons (of equal soliton mass) would start to behave similar to large molecules: after a collision all atoms of one kind would be measured either be on one side or on the other of the center of mass. Such cases are relevant for entanglement generation, and form the subject of the main paper.

\end{appendix}
%

\end{document}